\pdfoutput=1 

\documentclass[12pt]{article}
\usepackage[margin=3cm]{geometry}
\usepackage{amsfonts,amsmath,amssymb,amsthm}  
\usepackage{times}
\usepackage{graphicx}
\usepackage{color}
\usepackage{url}
\usepackage[bookmarks=false,colorlinks=true,urlcolor=blue]{hyperref}
\usepackage{cite}

\newtheorem{rem}{Remark}

\title{\sf Hypernets -- Good (G)news for Gnutella} 
\author{N. J. Gunther \\ {\small Performance Dynamics Research, Castro Valley, California} \\
{\small \url{www.perfdynamics.com}}
}

\normalsize
\date{February 16, 2002}

\begin{document}
\maketitle

\begin{abstract}
Criticism of Gnutella network scalability has rested on the bandwidth attributes
of the original interconnection topology: a Cayley tree. Trees, in general, are
known to have lower aggregate bandwidth than higher dimensional topologies e.g.,
hypercubes, meshes and tori. Gnutella was intended to support thousands to
millions of peers. Studies of interconnection topologies in the literature,
however, have focused on hardware implementations which are limited by cost to a
few thousand nodes. Since the Gnutella network is virtual, hyper-topologies are
relatively unfettered by such constraints. We present performance models for
several plausible hyper-topologies and compare their query throughput up to
millions of peers. The virtual hypercube and the virtual hypertorus are shown to
offer near linear scalability subject to the number of peer TCP/IP connections
that can be simultaneously kept open.
\end{abstract}

\section{Introduction}
The Gnutella network is a class of open source
virtual networks
known as \emph{Peer-to-Peer} or P2P networks.  Compared to the more ubiquitous
client-server distributed architectures, every P2P node (or \emph{servant}) can
act as both a client and a server. Many client-server applications such as, 
commercial databases, have multiple clients (users) accessing a centralized
server (see~\cite{njg00} Chap. 8). Conversely, P2P network applications
are usually completely decentralized.

Finding applications that can make efficient use of P2P is the current gating
factor for their widespread adoption. So far, P2P networks have been employed
for such applications as the Napster (\texttt{www.napster.com})
music file-sharing service, and the SETI@Home project
(\texttt{setiathome.ssl} \texttt{.berkeley.edu}), although those
implementations rely on a significant centralized server component.

\begin{rem}[Update 2012]
One of the best known working examples of a P2P-style computing system
is \href{Skype}{www.skype.com}, which allows millions of people to use
their PCs like a free telephone by forming its own gargantuan network
that supports the Voice Over Internet Protocol (VOIP). 

Other well-known
P2P architectures include Napster, Freenet, Limewire, Kazaa, and 
BitTorrent. 
Moreover, many of
these P2P architectures have progressed from file-sharing protocols
to a viable means for distributing applications such as games,
movies, and even software itself. For example,  
moving a 1.5 GB binary blob to countless servers is a non-trivial 
technical challenge. After exploring several solutions, Facebook.com came up 
with the idea of using 
\href{https://arstechnica.com/information-technology/2012/04/exclusive-a-behind-the-scenes-look-at-facebook-release-engineering/}{BitTorrent} for that purpose.
\end{rem}

The initial release of Gnutella in 2000 led to the perception that the
intrinsic architecture may not be capable of scaling to meet the sharing
demands of millions of anticipated
\footnote{
In 2001, the size of the Napster network was 160,000 simultaneous
users, down from a peak of 1.6 million reported by Webnoize in February, 2001
} users. Similar concerns about scalability
have arisen in the context of hypergrowth traffic impinging on popular
e-commerce Web sites~\cite{njg01}. Based on measurements of popular
queries,~\cite{caching} proposed that Gnutella scaling problems could be
ameliorated through the implementation of appropriate caching strategies.
Measurements by \cite{adar} indicated that there were more readers than writers
involved in file sharing. They suggested that such a ``free ride'' could lead to
higher than expected load on the P2P network thereby degrading its performance
as well as increasing its vulnerability to fragmentation.

A mathematical analysis by~\cite{ritter} (one of the original developers of
Napster) presented a detailed numerical argument demonstrating that the Gnutella
network could not scale to the capacity of the competitor \footnote{At
the height of the media attention, Napster's legal problems drove some 50,000
users per day over to Gnutella such that peers connected by 56 Kbps phone lines
caused the P2P network to fragment into disconnected ``islands'' of about 200
peers.} 
Napster network. Essentially, that model showed that the Gnutella network is
severely bandwidth limited long before the P2P population reaches a million
peers. In each of these previous studies, the conclusions have overlooked the
intrinsic bandwidth limits of the underlying topology~\cite{topos} in the
Gnutella network: a Cayley tree~\cite{cayley}. (See section ~\ref{sec:trees}
for the definition)

Trees are known to have lower aggregate bandwidth than higher dimensional
topologies e.g., hypercubes and hypertori. Studies of interconnection topologies
in the literature have tended to focus on hardware implementations (see
e.g.,~\cite{logp},~\cite{clusters}, ~\cite{parallel} and ~\cite{patterson})
which are generally limited by the cost of the chips and wires to a few thousand
nodes~\cite{njg02}. P2P networks, on the other hand, are intended to support
hundreds of thousands to millions of simultaneous peers and since they are
implemented in software, hyper-topologies are relatively unfettered \footnote{As
the SETI@Home project has demonstrated, 2.8 million desktops (and 10 PetaFLOPS)
can be harnessed for free.} by the economics hardware.

In this paper, we analyze the scalability of several alternative topologies and
compare their throughput up to 2-3 million peers. The virtual hypercube and the
virtual hypertorus offer near-linear scalable bandwidth subject to the number of
peer TCP/IP connections that can be simultaneously kept open. We
adopt the abbreviation \emph{hypernet} for these alternative topologies. The
assumptions about the distribution of peer activity are similar to those
employed by~\cite{ritter}. This is appropriate since our purpose is to rank the
relative performance of these hypernets rather than to predict their
absolute performance.

\section{Tree Topologies} \label{sec:trees}
In the subsequent discussion, the P2P network is treated as a graph i.e., a set
nodes or vertices connected by a set of edges or links. The nodes correspond to
network peers and the links to the links to network connections.

Because the tree structure of the Gnutella network has been such a hidden
determinant underlying the conclusions drawn in previous scalability studies, we
commence our performance comparisons by distinguishing clearly among the
relevant tree topologies. Topologically, all trees are planar and thus have  d =
2 spatial dimensionality.

\subsection{Binary Tree}
The binary tree is familiar in the computing context by virtue of its ubiquity
as a parsing and storage data structure~\cite{wirth}. There is a unique root
node which is connected only to two sibling nodes and each of those siblings is
connected to another pair of sibling nodes and so on. At each level (h) in the tree,
there are $2^{h}$ nodes. Therefore, the number of nodes grows as a binary
exponential. Because of its relatively sparse nodal density, the binary tree is
rarely employed as a bona fide interconnection network.

\subsection{Rooted Tree}
A rooted tree is simply the generalization of a binary tree in which
each node (other than the root) has a vertex of degree v.
The total number of nodes is the sum of a geometric series:
\begin{equation}
N_{bin}(h) = \frac{v^{h} - 1}{v - 1} ~\label{eqn:bin}
\end{equation}

\subsection{Cayley Tree}
A Cayley tree~\cite{cayley} has no root.  Recalling the binary tree,
what was the root of the parent binary tree now has
a link to an another binary sub-tree of height one less than the parent.
All nodes thus become tri-valent with v = 3 at every level.
More generally, for a v-valent tree, the total number of nodes is given by:
\begin{equation}
N_{cay}(h) = 1 + \sum~v ~(v - 1)^{h - 1}
\end{equation}
and therefore is denser than (~\ref{eqn:bin}).

This is the central formula used in the scalability analysis of~\cite{ritter}.
The network he analyzed is thus a Cayley tree with vertex degree
(v) corresponding to the number of open network connections per servant.
\cite{ritter} analyzed valences in the range $v = 4~\dots ~8$; the former value
being the default setting in the original Gnutella release, and the latter more
closely resembling the number of peers claimed for the contemporaneous Napster 
network.

\section{Hypernet Topologies} \label{sec:hypers}
An alternative to bandwidth-limited trees is a topology with
higher dimensionality. We examine the performance attributes of two hypernets
in particular: the binary hypercube and the hypertorus, each in 
d-dimensions.

\subsection{Hypercube}
In a boolean or binary hypercube each node forms the vertex of a d-dimensional
cube~\cite{hpcc}. The number of nodes is simply  $2^{d}$ and the degree of each
vertex (v) is equal to the dimensionality (d) of the network.
Hence, each node can be enumerated or addressed using a base-2 (binary) d-digit
number.

Moreover, since neighboring nodes differ in address by only 1 digit, sending a
message on the hypercube becomes a simple matter of shifting successive bits as
the binary address passes each node between source and destination.

In d = 3 dimensions the hypercube is simply a cube. Each vertex has degree v =
3, so there are $2^{3} = 8$ nodes. A 4-dimensional hypercube, can be visualized
as spatially translating a 3-cube such that the locus of its 4 vertices trace
out the additional connections.

\subsection{HyperTorus}
A d-dimensional hypertorus~\cite{hpcc} is a d-dimensional grid with each nodes
connected to a ring of nodes in each of the d orthogonal dimensions. The
hypertorus reduces to the binary hypercube when there are only 2 nodes in each ring.

The simplest visualization is, once again, in 3-dimensions. A 2-dimensional grid
is first wrapped about one axis such the edges join to form a tube. The tube is
wrapped about the orthogonal axis to form a ring such that the open ends of the
tube become joined. The result is a 3-torus, otherwise known as a donut.

All of these topologies fall into a class known as single stage networks and are
relatively easy to implement in software. The more exotic topologies, such as
cube-connected cycles, butterflies and other multistage~\cite{parallel} networks
are not considered here because they are likely to be more difficult to implement.

\section{Performance Metrics} \label{sec:metrics}

\subsection{Network Diameter ($\delta$)}
The notion of a network diameter is analogous to the diameter for a circle.
There, it is the maximum chordal length between two points on the circumference.
For a network, it is the maximum number of communication links that must be
traversed to send a message to any node along the shortest path. It represents
a lower bound on the latency to propagate messages throughout the entire network.
\begin{table}[!hbtp] 
\begin{center}
\begin{tabular}{|l|c|}
\hline
\textbf{Topology}	& \textbf{$\delta$} \\
\hline \hline
Tree		         & 	$2 h$		\\
Hypercube	& 	$d$		\\
Torus		& 	$\frac{d N^{1/d}}{4}$		\\
\hline
\end{tabular}
\caption{Network diameters.} \label{tab:diameters}
\end{center}
\end{table}
In 1997 the Web was estimated to comprise more than half a million
sites~\cite{gray}. By 2001, it was estimated~\cite{oclc} to have grown to 3.1
million publicly accessible sites.

The diameter of the Web has been estimated~\cite{diameter} to be about 20 hops.
If the Web is modelled as a Cayley tree, its height would be half the
diameter i.e., $h = \delta /2 = 10$ hops. A vertex degree of 5 (connections per node)
would contain just under half a million nodes while a vertex degree of 6 would
contain nearly 3 million (2,929,687) nodes.

\subsection{Total Nodes (N)}
The total number of peer nodes in the P2P network.
For a binary tree:
\begin{equation}
N(h) =  \sum_{k = 1}^{h} ~2^{k - 1} ~\label{eqn:nodes}
\end{equation}
For a d-dimensional binary hypercube the number of nodes is $2^{d}$.

\subsection{Path Length}
The path length is the maximal distance between a leaf node and the root. 
For a tree, it is half
the diameter. The path length corresponds the peer horizon used by~\cite{ritter}
in his analysis. A better measure of network latency is the average number of
hops (H), which we shall define shortly.

\subsection{Internal Path Length (P)}
The internal path length is the total number of paths between all nodes.
For a binary tree of depth h, the total number of paths is: 
\begin{equation}
P(h) =  \sum_{k = 1}^{h} ~k ~N(k) ~\label{eqn:path}
\end{equation}

\subsection{Average Number of Hops (H)}
Since the network diameter is a maximal distance, it tends to overestimate
message latency. A better measure is the average number of hops between
source and destination.
This quantity is found by dividing the internal path length in (\ref{eqn:path}) 
by the total number of nodes in (\ref{eqn:nodes}) 
 \begin{equation}
H =  \frac{P}{N} ~\label{eqn:hops}
\end{equation}
It corresponds to the average number of network hops traversed by a P2P query.

\subsection{Number of Network Links (L)}
This is a measure of the number of physical network links.
\begin{table}[!hbtp] 
\begin{center}
\begin{tabular}{|l|c|}
\hline
\textbf{Topology}	& \textbf{L} \\
\hline \hline
Tree			& 	$N_{tree}$		\\
Hypercube	& 	$\frac{dN_{cube}}{2}$ \\
Torus		& 	$dN_{torus}$		\\
\hline
\end{tabular}
\caption{Network links.} \label{tab:links}
\end{center}
\end{table}
As shown in Table~\ref{tab:links}, L scales with the number of physical nodes
(N) for the topologies we consider.

\subsection{Network Demand ($D_{link}$)}
The transit frequency across a link  $f_{link}$ 
is a measure of the average query size per link. Under the assumption of
uniform message routing, it can be defined as:
\begin{equation}
f_{link} = \frac{H}{L}
\end{equation}
If the latency across a link is denoted by $S_{link}$,
then the total service demand~\cite{njg00} is:
\begin{equation}
D_{link} = f_{link}~S_{link} ~\label{eqn:netD}
\end{equation}
For simplicity, and without loss of generality, we normalize the network demand
to unit periods ($S_{link} ~=~ 1$).

\subsection{Peer Demand ($D_{peer}$)}
Similarly, for node latency $S_{peer}$.
Under the assumption of uniform message routing:
\begin{equation}
f_{peers} = \frac{1}{N}
\end{equation}
and the total peer service demand is:
\begin{equation}
D_{peers} = \frac{S_{peer}}{N} ~\label{eqn:peerD}
\end{equation}
Again, we normalize the peer demand to unit periods ($S_{peer} ~=~ 1$) in the
subsequent discussion.

\subsection{Bandwidth (X)}
It follows from Little's law, $U = X D$ (See e.g.,~\cite{njg00} p. 44)
that when any node in the network reaches saturation ($U = 1$) 
the maximum in the system throughput is determined by:
\begin{equation}
X_{max} =  \frac{1}{Max[D_{peers}, D_{link1}, D_{link2}, ...]} ~\label{eqn:thruput}
\end{equation}
The node with the longest service demand $D_{max}$ is the system bottleneck. The
service demand at the bottleneck therefore determines the maximum system
throughput.

With these metrics defined, we are in a position to compare the asymptotic
performance of each of the topologies described in sections ~\ref{sec:trees} and
~\ref{sec:hypers}.

\section{Relative Bandwidth}
Since we are interested in network scalability up to a few million peers,
it is sufficient to base the comparison on the asymptotic network throughput
defined in (~\ref{eqn:thruput}). In particular, we will rank the above hypernets
according to their relative maximal bandwidth,
\begin{equation}
X_{relative} = X_{max}(N) / N \label{eqn:xrelative}
\end{equation}
where N is the number of peers in the horizon (Table \ref{tab:ranking}
at the end of this section).
$X_{relative} = 1.0$ corresponds to linear scalability since
$X_{max} = N$ in (\ref{eqn:xrelative}).

In several respects our approach is similar to that taken by~\cite{logp} for
their LogP model of assessing parallel hardware performance. In both approaches,
the respective network topology enters into the performance model via
the network demand defined in (~\ref{eqn:netD} and ~\ref{eqn:peerD}).

\subsection{Cayley Trees}
\begin{figure}[!ht]
\begin{center} 
\includegraphics[scale=0.5]{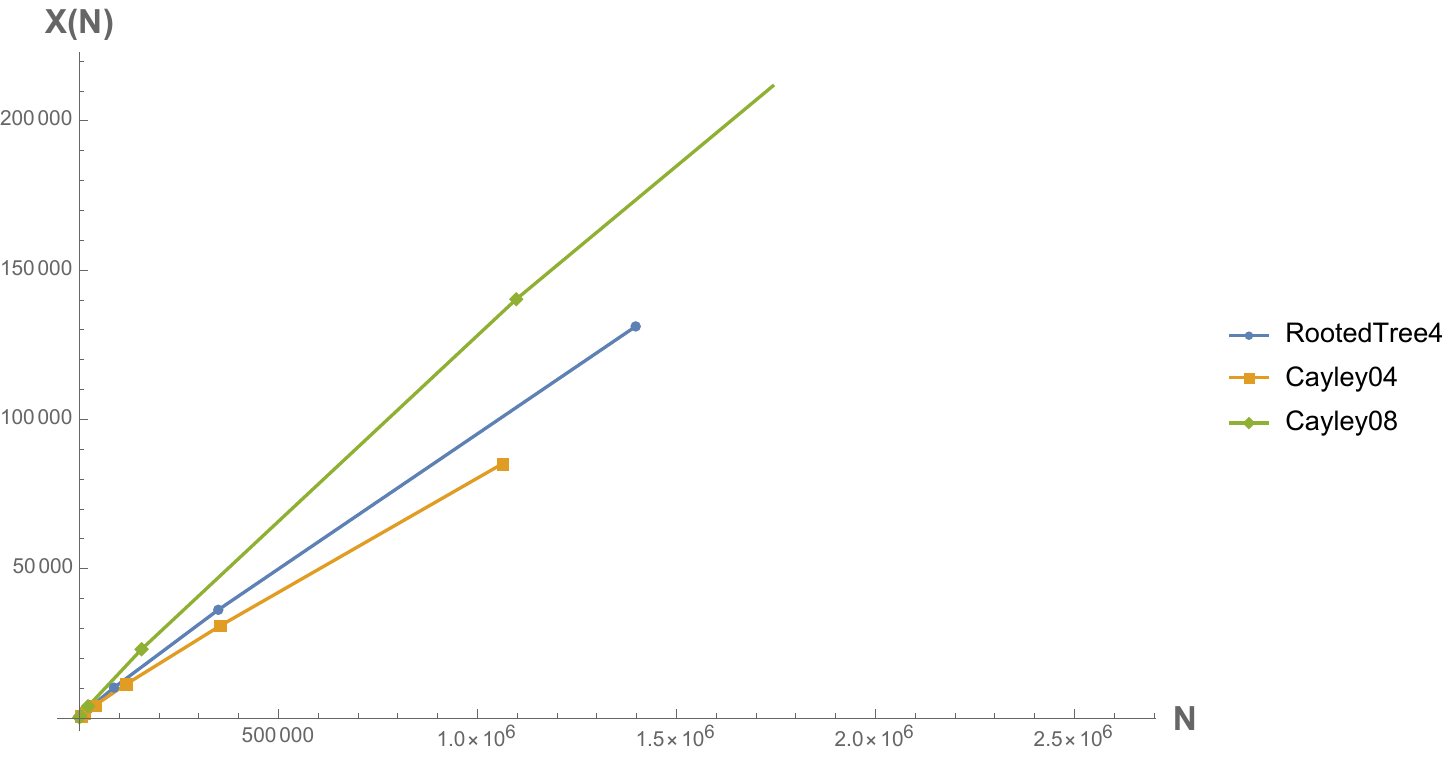} 
\caption{Comparative bandwidth of 2-dimensional trees} ~\label{fig:bincay}
\end{center}
\end{figure}

First, we consider the relative performance of tree topologies. Fig.~\ref{fig:bincay}
shows the normalized bandwidths of a 4-th degree rooted tree, a 4-valent Cayley
tree and an 8-valent Cayley tree.

The 4-valent Cayley tree represents the default peer connectivity in the original
release of Gnutella. Similarly, the 8-valent Cayley tree corresponds to Ritter's
comparison with Napster scalability.
The curves in Fig.~\ref{fig:bincay} terminate at different peer populations
because the population is an integral multiple which is dramatically affected by
the vertex degree and the height of the tree.

We see immediately that the 8-valent Cayley tree has the greatest bandwidth up
through 2 million peers. The 4-valent Cayley tree has the lowest bandwidth; even
lower than the rooted tree. This follows from the fact that at its root the
4-tree has the same connectivity as the 4-Cayley tree but all its descendents
have vertices of 5 degrees. Even for the 8-Cayley, at 2 million peers the
bandwidth is less than one quarter of linear scalability.

\subsection{Cubes and Trees }
\begin{figure}[!ht]
\begin{center} 
\includegraphics[scale=0.5]{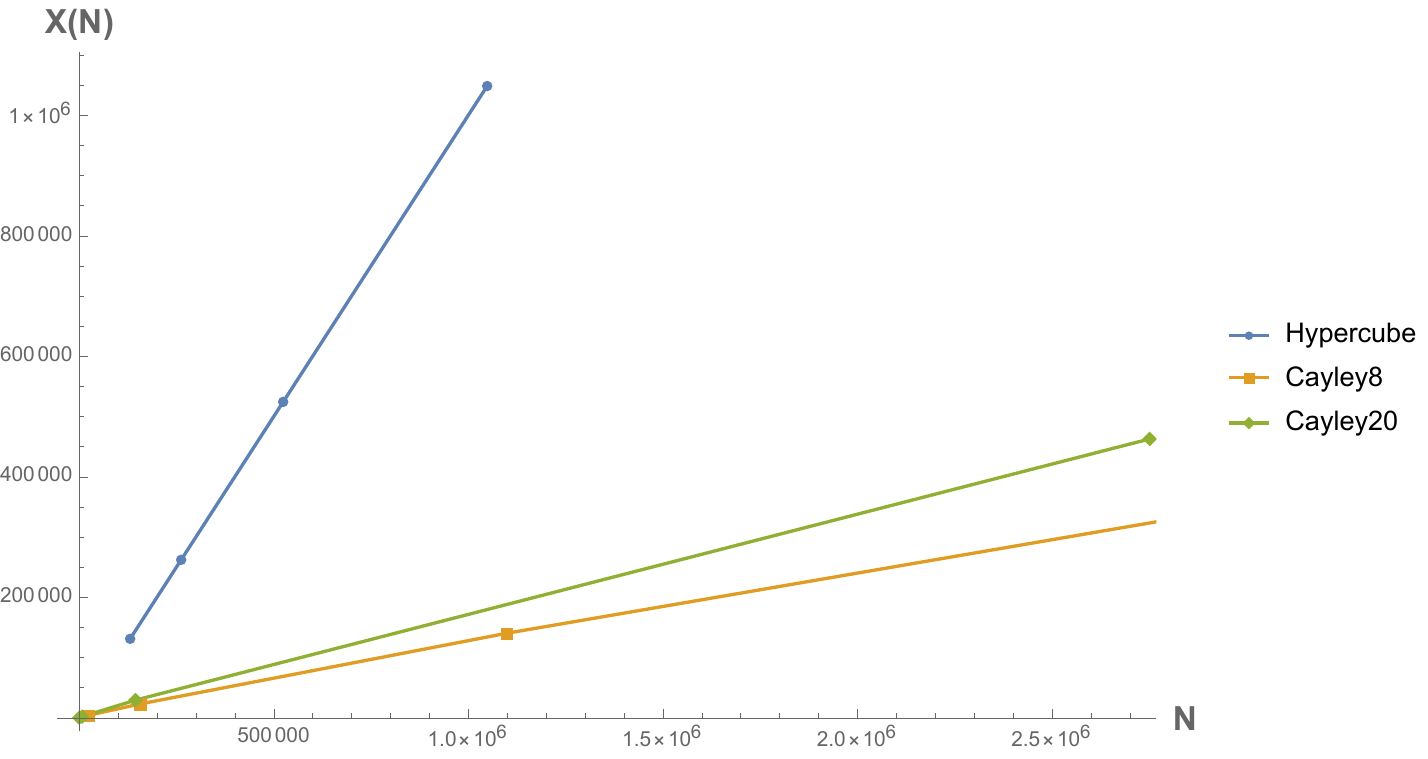} 
\caption{Comparative bandwidth of Cayley trees with the hypercube} ~\label{fig:cubtree}
\end{center}
\end{figure}

We next consider the relative performance of high degree trees and hypercubes. 
In particular, Fig.~\ref{fig:cubtree} shows the normalized bandwidths for an
8-Cayley (the best throughput of the trees considered in Fig.~\ref{fig:bincay}),
a 20-Cayley, and a binary hypercube. The d-dimensional hypercube clearly exhibits 
superior scalability.

\subsection{Cubes and Tori}
\begin{figure}[!hb]
\begin{center} 
\includegraphics[scale=0.5]{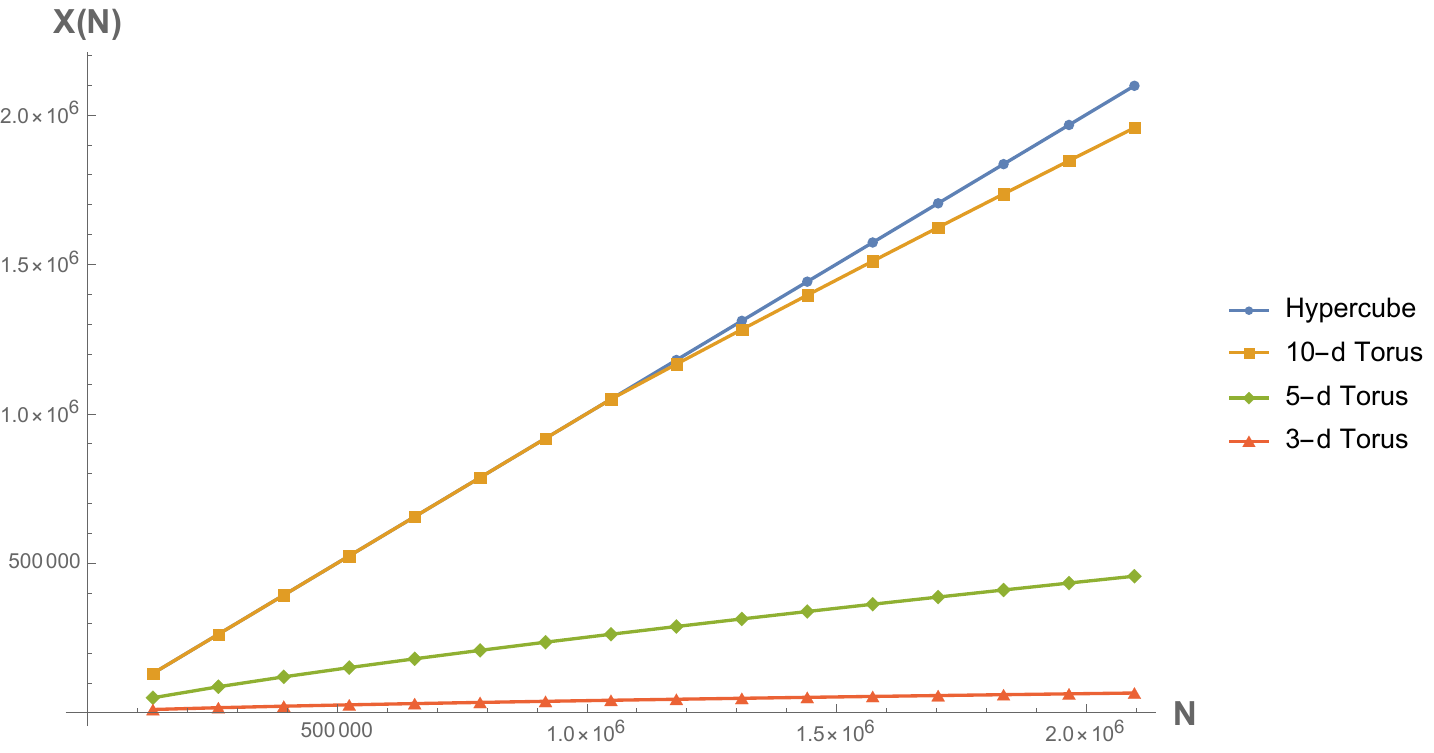} 
\caption{Comparative bandwidth of hypertori with the hypercube} ~\label{fig:cubtor}
\end{center}
\end{figure}

Of these high-order topologies, the binary hypercube offers linearly scalable
bandwidth beyond one million active peers (Fig.~\ref{fig:cubtor}). The
10-dimensional hypertorus has comparable scalability up to one million peers but
degrades beyond that point.
The 3-dimensional hypertorus is also shown for comparison since that topology
has been used in large-scale hardware implementations up to several hundred
nodes per cluster (e.g., the Tandem Himalya).

\subsection{Ranked Performance}
The main results of our analysis are summarized in Table~\ref{tab:ranking} which
shows each of the topologies ranked by their relative bandwidth as defined in
(\ref{eqn:xrelative}).
\begin{table}[!hbtp] 
\begin{center}
\begin{tabular}{|r|c|c|c|c|}
\hline
\textbf{Network} & \textbf{Connections} & \textbf{Hops to} & \textbf{Peers x $10^{6}$} & \textbf{Relative (\%)} \\
\textbf{Topology} & \textbf{per Peer}  & \textbf{Horizon} & \textbf{in Horizon}      & \textbf{Bandwidth} \\
\hline \hline
20-Cube		    & 	20  &	10	& 	2.1 	& 	100 \\
10-Torus		& 	20  &	11  & 	2.1 	& 	93 \\
5-Torus			& 	10  &	23  & 	2.1 	& 	22 \\
20-Cayley		& 	20	&	6	& 	2.8 	& 	16 \\
\textcolor{red}{8-Cayley} & 	\textcolor{red}{8}	& \textcolor{red}{8}	& \textcolor{red}{1.1} 	& 	\textcolor{red}{13}  \\ 
4-Tree			& 	4	&	11  & 	1.4 	& 	12 \\
3-Torus			& 	6  &	96  & 	2.1 	& 	10  \\  
\textcolor{red}{4-Cayley} & 	\textcolor{red}{4}	&	\textcolor{red}{13}	& 	\textcolor{red}{1.1} & 	\textcolor{red}{8} \\
\hline
\end{tabular}
\caption{Topologies ranked by maximal relative bandwidth.} \label{tab:ranking}
\end{center}
\end{table}

The 20-dimensional hypercube outranks all other contenders on the basis of query
throughput.  For an horizon containing 2 million peers, each servant
must maintain 20 open connections, on average. This is well within the capacity
limits of most TCP/IP implementations~\cite{stevens}.

The 10-dimensional hypertorus is comparable to the 20-hypercube in bandwidth up to 
an horizon of 1 million peers but falls off by almost 10\% at 2 million peers. The
10-torus is also arguably a more difficult topology to implement.

The 20-valent Cayley tree is included since the number of connections per peer is
the same as that for the 20-cube and the 10-torus. An horizon of 6 hops was
used for comparison because the peer population is only 144,801 nodes at 5 hops.
Similarly for 8-Cayley, a 9 hop horizon would contain 7.7 million peers.
These large increments are a direct consequence of the high vertex degree per node.

The 4-Cayley (modeling early Gnutella) and 8-Cayley (modeling the Napster
population) show relatively poor scalability at 1 million peers. Even doubling
the number of connections per peer produces slightly better than 50\%
improvement in throughput. This confirms the conclusions reached in
~\cite{ritter} and, moreover, supports our proposal to consider hypernet
topologies.

\section{Conclusions}
Previous studies of Gnutella scalability have tended to overlook the intrinsic
bandwidth limits of the underlying tree topology. The most thorough and accurate
of these studies is that presented in ~\cite{ritter}. Unfortunately, his
analysis could be accused of straining at a gnat. As a viable candidate for
massively scalable bandwidth, our analysis demonstrates that trees are dead.

Conversely, by going to higher dimensional virtual networks (and the
hypercube in particular) near linear scalability can be achieved for populations
on the order of several million peers each with only 20 open connections.
According to section \ref{sec:metrics}, this level of scalability would already
match the number of nodes present in the entire Web.

The dominant constraint for hardware implementations of high-dimensional networks
is the cost of the physical wires on the interconnect backplane. Since the
hypernets discussed here would be implemented in software, no such constraints 
would prevent reaching the desired level of scalability. In this sense, we see
hypernets as offering good (g)news for Gnutella scalability.

\let\oldbibliography\thebibliography
\renewcommand{\thebibliography}[1]{%
  \oldbibliography{#1}%
  \setlength{\itemsep}{0pt}%
}


\begin{thebibliography}{99}
\bibitem{adar}
	Adar E. and Huberman, B. A. 
	``Free Riding on Gnutella,'' 
	\url{http://www.firstmonday.dk/issues/issue5\_10/adar/index.html},
	October 2000.
\bibitem{parallel}
	Almasi, G. S., and Gottlieb, A. 
	\emph{Highly Parallel Computing,} 
	Benjamin-Cummings 1994.
\bibitem{clusters}
	Buyya, R. (Ed.) 
	\emph{High Performance Cluster Computing. Vol. 1, 
	Architectures and Systems,} 
	Prentice-Hall 1999.
\bibitem{logp}
	Culler, D. E., Karp, R. M., Patterson, D., Sahay, A., Santos, E. E., 
	Schauser, K. E., Subramonian, R., and Eicken, T. 
	``LogP: A Practical Model of Parallel Computation,'' 
	Comm. ACM, \textbf{39} (11), 79 -- 85, November 1996.
\bibitem{source}
	\url{http://core.limewire.org/servlets/ProjectHome} 
	Current development projects.
\bibitem{gray}
	Gray, M. K. 
	``Web Growth Summary,'' 
	\url{http://www.mit.edu/people/mkgray/net/web-growth-summary.html}
\bibitem{njg00}
	Gunther, N. J. 
	\emph{The Practical Performance Analyst,}  
	iUniverse.com Inc. 2000. 
	The relevant sections can be read online at 
	\url{http://books.iuniverse.com/viewbooks.asp?isbn=059512674X1\&page=fm5}
\bibitem{njg01}
	Gunther, N. J. 
	``Performance and Scalability Models for a Hypergrowth e-Commerce Web Site,''
	in
	\emph{Performance Engineering: State of the Art and Current Trends,}   
	(Eds.) Dumke, R., Rautenstrauch, C., Schmietendorf, A., Scholz, A., 
	\# 2047. Heidelberg: Springer-Verlag 2001.
\bibitem{njg02}
	\emph{Scalable Server Performance and Capacity Planning,} 
	UCLA Course 819.328 
	\url{http://www.perfdynamics.com/ucla.html},  
	March 2002.
\bibitem{hpcc}
	\emph{High Performance Computing and Communications Glossary}, 
	\url{http://nhse.npac.syr.edu/hpccgloss/hpccgloss.html}
\bibitem{topos}
	Minar, N. 
	``Distributed Systems Topologies,''
	\url{http://www.openp2p.com/pub/a/p2p/2002/01/08/p2p\_topologies\_pt2.html}, January 2002.
\bibitem{oclc}
	Online Computer Library Center report  
	\url{http://wcp.oclc.org/stats/size.html}
\bibitem{patterson}
	Hennessy, J. L., and Patterson, D. A. 
	\emph{Computer Architecture: A Quantitative Approach, (2nd Edition)}, 
	Morgan Kaufmann 1996.
\bibitem{cayley}
	Rains, E. M. and Sloane, N. J. A. 
	``On Cayley's Enumeration of Alkanes (or 4-Valent Trees),'' 
	\url{http://www.research.att.com/~njas/sequences/JIS/cayley.html} 
	\emph{Journal of Integer Sequences,} January 1999.
\bibitem{diameter}
	Reka, A., Hawoong, J., and Barabasi, A-L. 
	``The Diameter of the World Wide Web.'' 
	Nature \textbf{401} 130-131, 1999.
\bibitem{ritter}
	Ritter, J. 
	``Why Gnutella Can't Scale. No, Really.'' 
	\url{http://www.darkridge.com/~jpr5/doc/gnutella.html}, 
	Slashdotted on January 2002.
\bibitem{caching}
	Sripanidkulchai, K. 
	``The popularity of Gnutella queries and its implications on scalability,'' 
	\url{http://www-2.cs.cmu.edu/~kunwadee/research/p2p/gnutella.html} 
	March 2001.
\bibitem{stevens}
	Stevens, W. R. 
	\emph{UNIX Network Programming}, 
	Prentice Hall 1990.
\bibitem{wirth}
	Wirth, N.
	\emph{Algorithms + Data Structures = Programs},   
	 Prentice Hall 1976.
\end{thebibliography}
\end{document}